\documentclass[12pt,longnamesfirst,flushrt,preprint]{aastex} 

\newcommand{\eq}{\begin{equation}}
\newcommand{\feq}{\end{equation}}
\newcommand{\eqn}{\begin{eqnarray}}
\newcommand{\feqn}{\end{eqnarray}}

\shorttitle{Alfvenic heating of magnetic funnels}
\shortauthors{Vasconcelos, M.J., Jatenco-Pereira, V. & Opher, R.}

\begin{document}

\title{The role of damped Alfv\'en waves on magnetospheric accretion 
models of young stars}

\author{M. J. Vasconcelos}
\affil{Departamento de Ci\^encias Exatas e Tecnol\'ogicas, Universidade Estadual de Santa Cruz \\
Rodovia Ilh\'eus - Itabuna, km. 16, 45650-000 Ilh\'eus, BA, BRAZIL}
\email{mjvasc@uesc.br}

\and

\author{ V. Jatenco-Pereira and R. Opher}
\affil{Instituto de Astronomia, Geof\'{\i}sica e Ci\^encias Atmosf\'ericas \\
Universidade de S\~{a}o Paulo \\
Av. Miguel St\'efano 4200, 04301-904 S\~{a}o Paulo, SP, BRAZIL}

\begin{abstract}

We examine the role of Alfv\'en wave damping in heating the plasma in
the magnetic funnels of magnetospheric accretion models of young stars.
We study four different damping mechanisms of the Alfv\'en waves:
nonlinear, turbulent, viscous-resistive and collisional. Two different
possible origins for the Alfv\'en waves are discussed: 1) Alfv\'en
waves generated at the surface of the star by the shock produced by the
infalling matter; and 2) Alfv\'en waves generated locally in the funnel by
the Kelvin-Helmholtz instability. We find that, in general, the damping
lengths are smaller than the tube length. Since thermal conduction in
the tube is not efficient, Alfv\'en waves generated only at the star's
surface cannot heat the tube to the temperatures necessary to fit the
observations.  Only for very low frequency Alfv\'en waves $\sim 10^{-5}$
the ion cyclotron frequency, is the viscous-resistive damping length
greater than the tube length. In this case, the Alfv\'en waves produced at
the surface of the star are able to heat the whole tube. Otherwise, local
production of Alfv\'en waves is required to explain the observations. The
turbulence level is calculated for different frequencies for optically
thin and thick media. We find that turbulent velocities varies greatly
for different damping mechanisms, reaching $\sim 100$ km s$^{-1}$ for
the collisional damping of small frequency waves.

\end{abstract}

\keywords{accretion, accretion disks --- magnetic fields --- 
stars: pre-main-sequence --- turbulence --- waves}

\section{Introduction}

The current picture of Classical T Tauri stars (CTTSs), the so called
class II objects \citep{lada87}, consists of a central young star,
surrounded by a geometrically thin disk. Instead of a classical
boundary layer connecting the disk directly to the star, as proposed
by \citet{lynden...74}, the disk is disrupted by the star's magnetic
field at a given radius. Accretion flow for smaller radii follows the
star's magnetic field lines until it impacts on the stellar surface
\citep{ghosh...79a,ghosh...79b,konigl91}. This ``magnetospheric accretion
model" explains observational signatures seen in some CTTSs as, for
example, the excess of optical and ultraviolet continuum flux (veiling)
and redshift absorption features in the emission line profiles (inverse
P Cygni profiles) \citep*[hereafter HHC]{muz..98,hart..94}. Low rotation
rates, inferred from observations for CTTSs, are hard to explain by the
classical boundary layer model. However, they can be explained in the
magnetospheric accretion model.

O\-ri\-gi\-na\-lly suggested by \citet{ghosh...79a,ghosh...79b} for
neu\-tron stars, the mag\-ne\-tos\-phe\-ric ac\-cre\-tion mo\-del was
pro\-po\-sed to ex\-plain the hot s\-pots ob\-ser\-ved in DF Tau\-ri
by \citet{bertout...88}. \citet{camez90} and \citet{konigl91} applied
the original Ghosh-Lamb model to CTTSs. It is assumed, in this model,
that the star has a dipole magnetic field.  The basic idea behind this
mechanism is that a sufficiently strong field can halt disk accretion
at a given radius. At this radius, the magnetic pressure needs to be
equal to the ram accretion pressure. For protostellar accretion disks,
magnetic fields on the order of 1kG at the star's surface are sufficient
to disrupt the disk. Magnetic fields of this magnitude are inferred from
observations \citep{johns-krull...99,guenther...99}.  In order to follow
the magnetic field lines, the gas needs to be coupled to the magnetic
field. For CTTSs disks, this coupling is achieved if the temperatures at
the truncation radius ($R \leq 0.1$ AU) are greater than $10^3$ K since
collisional ionization of metal atoms is then effective \citep{ume...88}.
The truncation radius, $R_{trun}$, must be smaller than the co-rotation
radius, $R_{co}$, in order for accretion to proceed. Contrary to
observational evidences, for the case $R_{trun} < R_{co}$, the star
should spin up.  Wind models that possess open magnetic field lines
have been proposed to explain the observed low rotation rates. However,
the exact position (or region) of the truncation radius is still under
debate \citep{camez90,shu..94, hart98}. These open field lines carry
off mass and angular momentum. Several CTTSs do, in fact, show P Cygni
profiles in many lines, as well as in forbidden-line emission, indicating
the presence of outflows. Some CTTSs are known to have jets \citep[e.g.,
DG Tau, see][]{bacci..00}.

\citeauthor{hart..94} and more recently \citet{muz..98} calculated
magnetospheric accretion models, solving radiative transfer equations
in the Sobolev approximation, in order to obtain theoretical Balmer
line profiles. Profiles which are in reasonable agreement with observations 
were obtained. The calculated line profiles are slightly asymmetric,
with blueshifted centroids, sometimes showing redshifted absorption
components. However, \citet{alencar..00}, analyzing spectra from 30 CTTSs,
noted that only 20\% of their sample showed inverse P Cygni profiles
in the studied lines. They argued that winds, turbulence and rotation
of the central star must be taken into account to fully explain the
observations. \citet{basri90} previously suggested that turbulence
may be important in the formation of line profiles. \cite{edw..94}
suggested that Alfv\'en waves can be the source of this turbulence. Also,
\citet{johns..95}, when analyzing the spectra of SU Aur, found that a
turbulent velocity component at the base of a wind is necessary in order
to fit the observed spectra.

The major uncertainty in the radiative magnetospheric model calculations
is the temperature profile of the tube. \citet*[hereafter MA]{martin}
calculated the energy balance of the gas. He included heating by adiabatic
compression of the magnetic field lines, Balmer photoionization and
ambipolar diffusion. However, the temperatures that he obtained are very
low and cannot explain the observed line fluxes. Thus, an additional
heating mechanism must take place in the tube.

Combining the evidence for turbulence and the necessity of an additional
heating mechanism, we suggest that the heating by Alfv\'en waves
is important in the magnetic flux tubes of CTTSs. We first study
the possibility that the waves are generated at the star's surface
due to the shock produced by the accreting matter, as suggested by
\citet{scheur..88}. A model in which the waves are generated locally is
then studied, following the same approximation used in our previous paper
\citep*[ hereafter, Paper I]{vasc..00}. Independent of the generation
mechanism, we calculate the damping length for the waves, investigating
four different damping mechanisms: 1) nonlinear; 2) turbulent
(\citeauthor{vasc..00}); 3) collisional; and 4) viscous-resistive
\citep{oster61}. The associated heating rates and the degree of turbulence
are calculated. We use a theoretical temperature profile, varying from
$\sim 7500$ K to $\sim 8300$ K, which produces the observed line features.

Various damping mechanisms for Alfv\'en waves have been suggested in the
literature, such as Alfv\'en resonant heating of solar loops; wave damping
by phase mixing (also in the solar context); and cyclotron heating,
occurring as an Alfv\'en wave travels down a magnetic field gradient
until its frequency matches the decreasing ion cyclotron resonance
(magnetic beach). In addition to the more conventional collisional
and viscous-resistive Alfv\'en wave dampings, we concentrate on
nonlinear and turbulent damping, which our group has investigated
in: the solar wind \citep{vera..89a,vera..94}; protostellar winds
\citep{vera..89b}; late-type giant stars \citep{vera..89c}; Wolf-Rayet
stars \citep{santos..93a,santos..93b}; quasar clouds 
\citep{dede..93a,dede..96}; and extragalactic jets \citep{dede..93b}.

In \S (\ref{magneto}), we outline the magnetospheric accretion
model investigated. This accretion model is the same as that of
\citeauthor{hart..94}. The damping mechanisms used are discussed in \S
(\ref{dampmech}), where the heating rates, damping lengths and related
time-scales are evaluated for a reasonable set of parameters.  We discuss
the generation source for the waves and calculate the required degree
of turbulence in \S (\ref{sources}). Finally, the conclusions of this
work are summarized in \S (\ref{conc}).

\section{The magnetospheric accretion model} \label{magneto}

\begin{figure}[htb!]
\epsscale{0.8}
\plotone{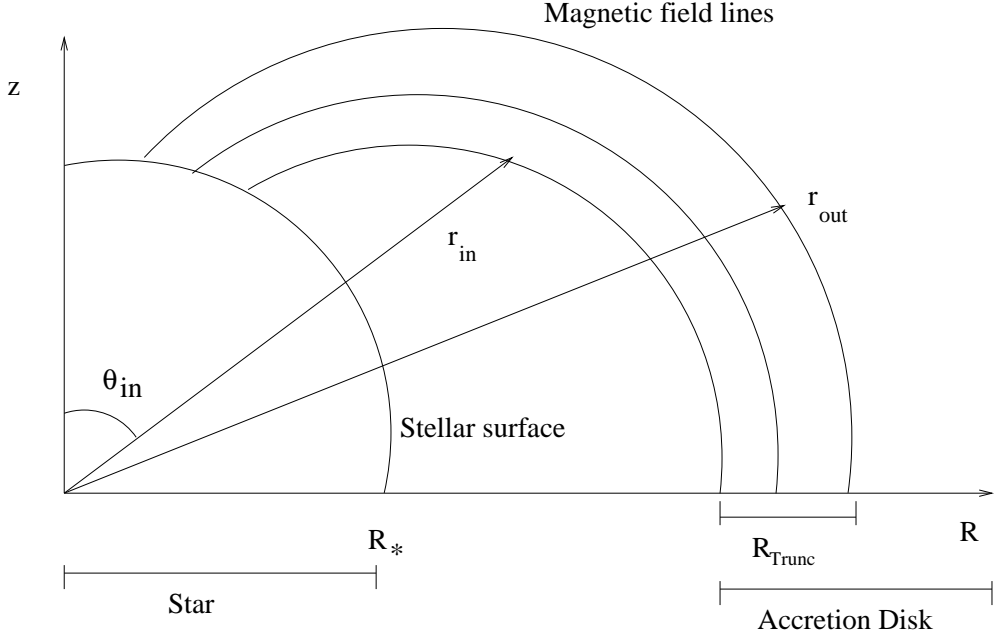}
\caption{{\small Sketch of the interface between the accretion disk and
the star. The axis shows the cylindrical coordinates $z$ (the vertical
direction) and $R$ (the projection of $r$ in the plane $z = 0$). The
star possesses a dipole magnetic field, which truncates the disk at the
radius $R_{trun}$. The angle between the z-direction and the position
vector $r$ is $\theta$. The inner and outer field lines are $r_{in}$
and $r_{out}$, respectively. (For more details of the geometry, see
\citeauthor{hart..94}).}
\label{fig1}} 
\end{figure}

The equations of the magnetospheric accretion model describe a column
of gas following the lines of a dipole magnetic field that connect
the disk to the star. In this paper, the equations used are the same
as those of \citeauthor{hart..94}. A cylindrical coordinate system
is adopted (see Fig. \ref{fig1})\footnote{For more details, see
\citeauthor{hart..94}.}. The streamlines are described by the equation

\eq \label{r}
r = r_m \sin^2 \theta,
\feq
where $\theta$ is the angle between the z-direction and $r$, the radial
distance from the center of the star. The constant $r_m$ is the value of
$r$ for $\theta = \pi/2$. We can also define a cylindrical radius $R$,
which is the projection of $r$ in the plane $z = 0$. In this paper,
all figures are shown as a function of the radius $R$.

The poloidal velocity, parallel to the field lines, is

\eq \label{vpvec}
{\bf v_p} = - v_p \left[\frac{3 \sin \theta \cos \theta {\bf \hat{R}} + 
(2 - 3 \sin^2 \theta){\bf \hat{z}}}{(4 - 3\sin^2 \theta)^{1/2}}\right],
\feq
where ${\bf \hat{R}}$ and ${\bf \hat{z}}$ are unit vectors in the $R$
and $z$ directions and

\eq \label{vp}
v_p = \left[\frac{2GM_{\ast}}{R_{\ast}}\left(\frac{R_{\ast}}{r} - 
\frac{R_{\ast}}{r_m}\right)\right]^{1/2},
\feq
where $M_{\ast}$ is the mass of the star and $R_{\ast}$ its radius.
The density of the fluid is

\eq \label{dens}
\rho = \frac{\dot{M}}{4\pi\left(\frac{1}{r_{in}} - \frac{1}{r_{out}}\right)}
\frac{r^{-5/2}}{\sqrt{2GM_{\ast}}}\sqrt{\frac{4 - 3\sin^2 \theta}
{\cos^2 \theta}},
\feq
where $\dot{M}$ is the accretion rate and $r_{in}$ and $r_{out}$ the
inner and outer field lines, respectively. A dipole magnetic field with 
intensity

\eq \label{B}
B = \frac{m \sqrt{4 - 3\sin^2 \theta}}{r^3},
\feq
where $m$ is the magnetic moment, is used.

\begin{figure}[htb!]
\plotone{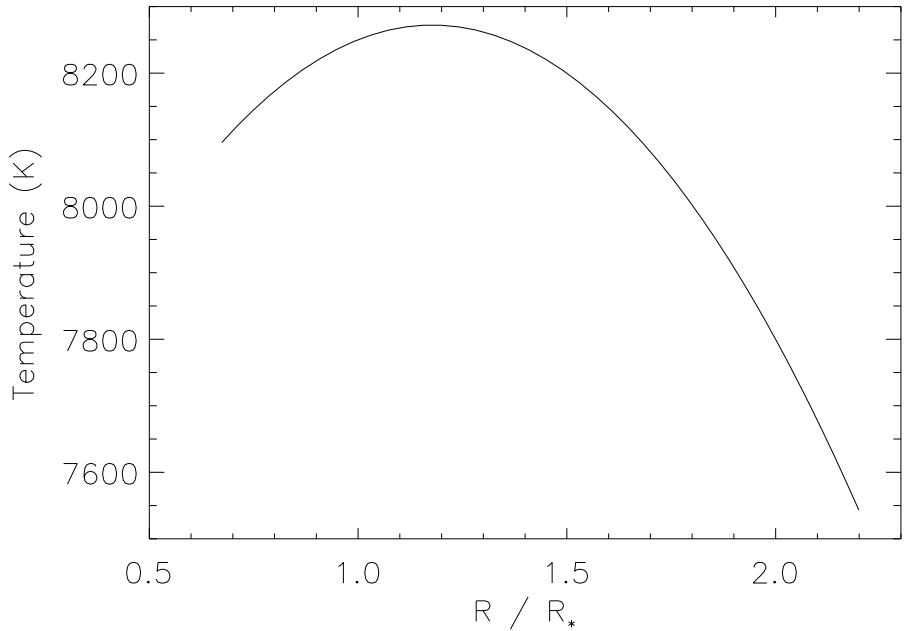}
\caption{{\small Temperature profile adopted in this work.
The profile fits the observed line fluxes \citep[e.g.][]{alencar..00}.
The temperature near the star does not vary greatly, dropping near the
disk. Since the Alfvenic heating does not have a strong dependency on
temperature, the exact profile is not very important.} \label{fig2}}
\end{figure}

It is not known what the exact temperature profile of the tube is.
\citeauthor{martin} calculated the thermal structure of the gas funnel,
but obtained values for the temperature that were too low to explain the
observations. \citeauthor{hart..94} adopted an ad hoc temperature profile
that was able to reproduce the observations. In the absence of a reliable
temperature estimate for the tube, we assume here a temperature profile
similar to that of \citeauthor{hart..94}, shown in Figure \ref{fig2}. The
mean temperature is $\sim 8000$ K. The temperature does not vary greatly
inside the tube, dropping near the disk. As we show in the following
sections, Alfvenic heating does not have a strong dependence on
temperature and, thus, the exact profile is not of any great importance.

In Table \ref{tab1}, we show the magnetospheric model parameters
adopted in our calculations, which are the same as those used by
\citeauthor{hart..94}.

\placetable{tab1}

\section{Damping mechanisms} \label{dampmech}

\subsection{The damping lengths}

When waves are generated in a non-ideal fluid, they are subject to
dissipation mechanisms so that the energy carried by the wave is transferred
to the fluid. We study four damping mechanisms in this work: 1)
nonlinear; 2) turbulent; 3) collisional; and 4) viscous-resistive. In
the nonlinear damping process, waves of large amplitudes interact and
generate compressive MHD waves which, in turn, decay into purely sonic
modes. The damping rate associated with this mechanism is

\eq \label{gamanl}
\Gamma_{nl} = \frac{1}{4} \sqrt{\frac{\pi}{2}} \xi \varpi
\left( \frac{c_s}{v_A} \right) 
\frac{\rho \langle \delta v^2 \rangle }{B^2/8\pi},
\feq
where $\xi$ is a constant that can assume a value between 5-10, $\varpi$
is the average angular frequency of the waves, $c_s$ is the sound
velocity, $v_A = B / \sqrt{4\pi\rho}$ is the Alfv\'en velocity, $\rho$ and
$B$ are the fluid density and magnetic field, respectively, and $\langle
\delta v^2 \rangle^{1/2}$ is the velocity perturbation, which is a
measure of the degree of turbulence of the system (\citeauthor{vasc..00}).

The turbulent damping mechanism involves an energy cascade from large to
small scales, in which microscopic processes dissipate the energy. The
associated damping length is

\eq \label{Ltur}
L_T = L_{corr} v_A \langle \delta v^2 \rangle^{-1/2},
\feq
where $L_{corr} (\propto B^{-1/2})$ is the correlation length
 of the turbulent vortices \citep{hollweg86}.

If the fluid is formed by two components, one ionized and the other
neutral, wave damping, driven by collisions between the positive ions
and the neutral particles \citep{oster61} can occur. The {\it e}-folding
length for the damping is

\eq \label{Lcol}
L_{coll} = \frac{v_A}{4\pi^2\nu^2}\frac{1 + \eta}{\eta \tau_n},
\feq
where $\nu$ is the average wave frequency, $\eta = \rho_H/\rho_i$ is the
density contrast between the neutral and ionized particles, and $\tau_n$,
the mean collision time.

For a hydrogen fluid, the mean collision time (in seconds) is 

\eq \label{tau}
\tau_n \approx 10^{12} T^{-1/2} n_{H}^{-1},
\feq
where $n_H$ is the neutral hydrogen particle density and $T$ is the
fluid temperature \citep{spi62}. The wave frequency can be written in
terms of the ion-cyclotron frequency, $\omega_i$,

\eq \label{nu}
\nu = \frac{\varpi}{2\pi} = \frac{F \omega_i}{2\pi} = \frac{FeB}{2\pi m_i c},
\feq
where $F$ is a free parameter ($F < 1$ corresponds to low-frequency
waves), $e$, $m_i$ and $c$ are the electron charge, ion (hydrogen)
mass and the velocity of light, respectively.

Substituting (\ref{tau}) and (\ref{nu}) in equation (\ref{Lcol})
we obtain,

\eq \label{Lcol2}
L_{coll} \approx v_A \left(\frac{m_i c}{FeB}\right)^2 
\frac{n_H \sqrt{T}}{10^{12}}\frac{1+ \eta}{\eta}.
\feq

Finally, we discuss the fourth damping mechanism which is related to
the viscosity $\mu$ and to the finite electric conductivity $(\sigma)$
of the gas. In this case, the {\it e}-folding length is,

\eq \label{Lsimu}
L_{\sigma, \mu} = \frac{v_{A}^{3}}{4\pi^2 \nu^2
\left(\frac{c^2}{4\pi\sigma} + \frac{\mu}{\rho}\right)}.
\feq
The electric resistivity $\eta_e$ for a partially ionized gas is 
\citep{spi62},

\eq \label{eeta}
\eta_e = \frac{c^2}{4 \pi \sigma} \approx 7.148 \times 10^{10} T^{-3/2}.
\feq
From \citet{spi62}, the viscosity for a gas composed mainly of
hydrogen is

\eq \label{mu}
\mu \approx 2 \times 10^{-15} T^{5/2} \;{\rm g \;cm^{-1} \; s^{-1}.}
\feq
Substituting equations (\ref{eeta}) and (\ref{mu}) in equation
(\ref{Lsimu}), we obtain

\eqn
L_{\sigma,\mu} & \approx & \left(\frac{m_H c}{F e B}\right)^2 \\ \nonumber
 & \times & \frac{v_A^3}{\left(7.148 \times 10^{10}T^{-3/2} + 
\frac{2 \times 10^{-15} T^{5/2}}{\rho}\right)},
\feqn
where the $F$ parameter is defined in equation (\ref{nu}).

\subsection{Heating rates}

As outlined in \citeauthor{vasc..00}, heating rates are given by

\eq \label{Ha}
H_A = \frac{\Phi_w}{L_A},
\feq
where $H_A$ is the heating rate associated with a generic damping
mechanism, $\Phi_w = \rho\langle \delta v^{2}\rangle v_{A}$ is the
wave flux and $L_A$ is the damping length. We derived the heating
rates for both the nonlinear and turbulent damping mechanisms in
\citeauthor{vasc..00}. They are, respectively:

\eq \label{Hnl}
H_{nl}=\frac{\sqrt{2}}{8}f^{4}F\frac{e\xi}{m_{i}c}
\left(\frac{\gamma \Re}{\bar{m}} \right)^{1/2}(\rho T)^{1/2}B^{2},
\feq
and

\eq \label{HT}
H_{T} \approx 2.985 \times 10^{-6} \frac{f^3 B^{7/2}}{\sqrt{\rho}},
\feq
where we used the same parametrization as that of \citeauthor{vasc..00},

\eq \label{dv2}
\langle \delta v^2 \rangle^{1/2} = f v_A,
\feq
and $f$ is a free parameter. As noted above, $\langle \delta v^2
\rangle^{1/2}$ is a measure of the degree of turbulence of the flow.
In equation (\ref{Hnl}), $\gamma = c_p/c_v$ is the ratio of the specific
heats, $\Re$ is the gas constant, and $\bar{m}$, the mean molecular weight.

Using equation (\ref{dv2}), the Alfv\'en wave flux can be written as,

\eq \label{fluxo}
\Phi_w = \rho f^2 v_{A}^{3} = f^2 \frac{B^3}{\sqrt{(4\pi)^3 \rho}}.
\feq

The collisional heating rate is given by,

\eq \label{Hcol}
H_{coll} \approx 1 \times 10^{12} \left(\frac{fFe}{m_H c}\right)^2 
\frac{B^4}{4 \pi (n_p + n_H) \sqrt{T}}.
\feq
Equation (\ref{Hcol}) is derived from equations (\ref{Lcol2}) and
(\ref{Ha}), using the definition of $\Phi_w$. In this equation, $n_p$
is the proton number density. Correspondingly, for the case when the
damping is due to both viscosity and the resistivity of the fluid, the
heating rate is

\eqn \label{Hvisres}
H_{\sigma, \mu} & \approx & \rho \left(\frac{fFeB}{m_H c}\right)^2 \\ \nonumber
 & \times & \left(7.2 \times 10^{10} T^{-3/2} + 
\frac{2 \times 10^{-15}T^{5/2}}{\rho}\right).
\feqn

\begin{figure}[htb!]
\plotone{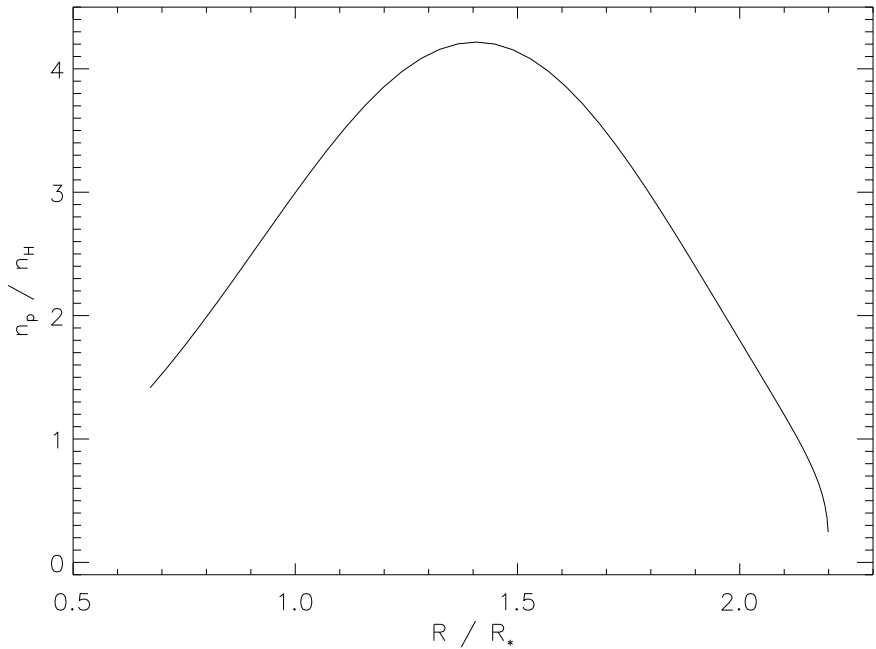}
\caption{{\small Ratio of protons to neutral atoms. We note that the
ratio drops near the disk. This is because the temperature decreases
there and the density increases. Near the star the ratio also drops,
but not as dramatically as near the disk.} \label{fig3}}
\end{figure}

\begin{figure}[htb!]
\epsscale{1.08}
\plotone{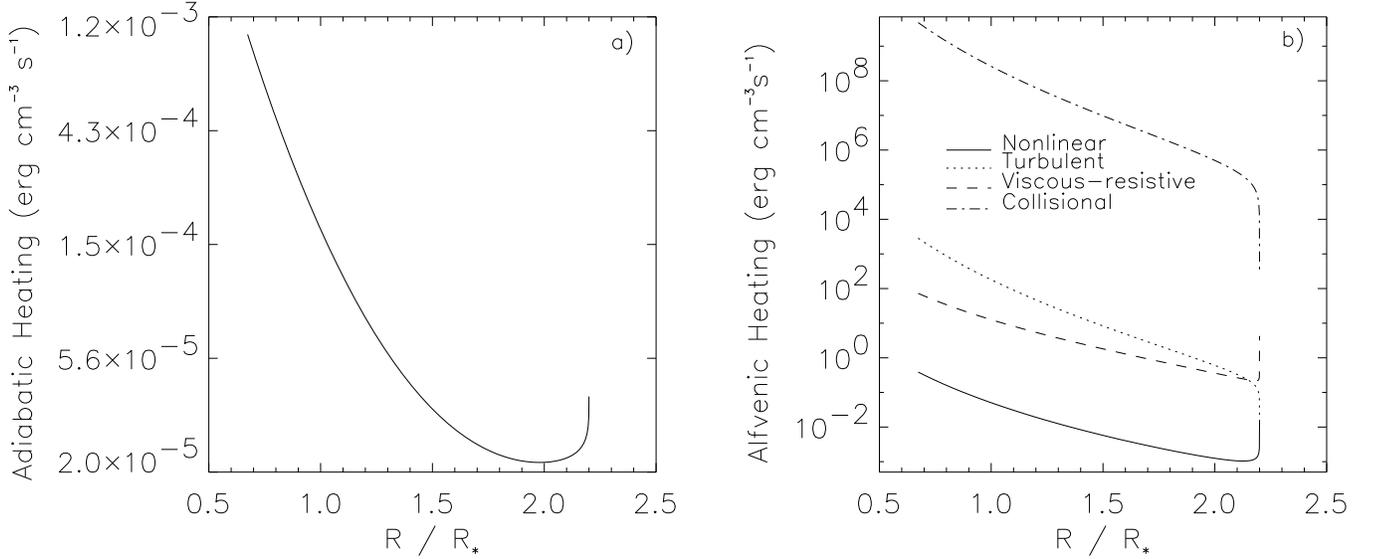}
\caption{{\small Heating rates (ergs ${\rm cm}^{-3}$ ${\rm s}^{-1}$):
4a) Adiabatic, calculated according to \citeauthor{martin}; 4b)
Alfvenic for nonlinear (solid line), turbulent (short-dashed line),
viscous-resistive (long-dashed line) and collisional (dash-dotted
line) damping mechanisms. We note that the heating rates vary from
$\sim 10^{-2}$ to $\sim 10^{11}$ erg ${\rm cm}^{-3}$ ${\rm s}^{-1}$.}
\label{fig4ab}} 
\end{figure}

\begin{figure}[htb!]
\plotone{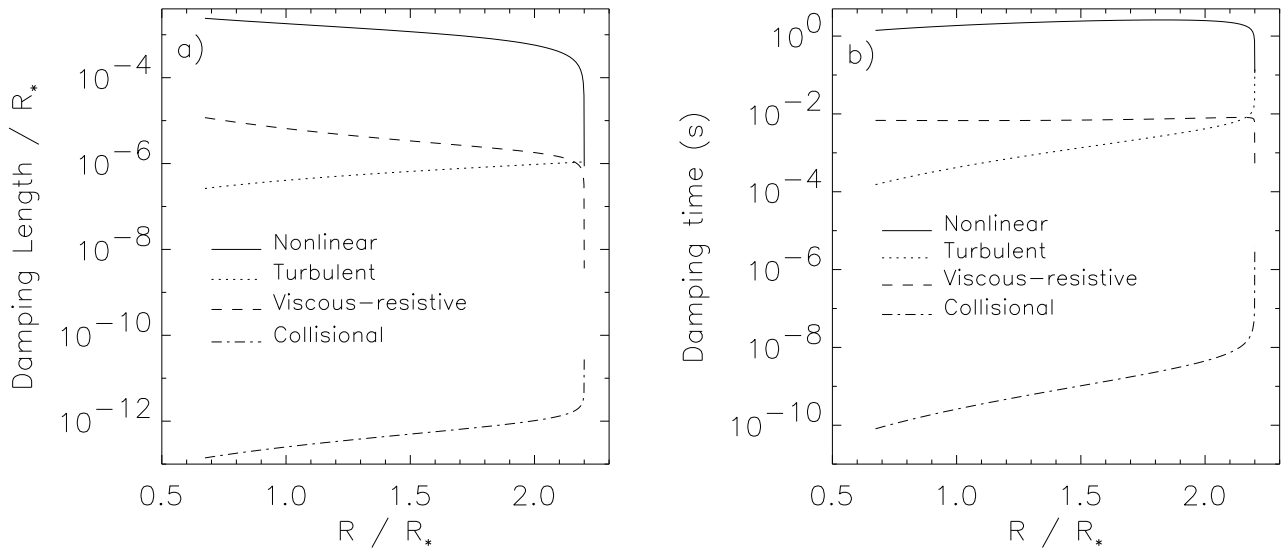}
\caption{{\small Damping lengths and time scales: 5a) Damping lengths
calculated for the mechanisms described in section \ref{dampmech}. The
lengths are divided by the stellar radius. We note that the lengths
are very small, compared to the stellar radius. Thus we can assume that
as soon as the waves are generated, they are rapidly damped. (For more
discussion, see text). 5b) Time-scales related to nonlinear (solid line),
turbulent (short-dashed line), viscous-resistive (long-dashed line) and
collisional (dash-dotted line) damping mechanisms.  The time-scales vary
from $\approx 10^{-10}$ s for the collisional mechanisms to $\approx 3$
s for the nonlinear mechanism.  Even the slowest time-scale is very
short, compared to the free-fall time-scale which is $\approx 9$ hr.}
\label{fig5ab}}
\end{figure}

In order to calculate the heating rates and damping lengths described
above, we set values for the variables, using the parameters of Table
\ref{tab1}. We use the following values: $f=0.004$ (justified in later
sections), $F=0.1$, $\xi = 5$, $m_i = m_H = 1.673 \times 10^{-24}$ g,
$\gamma = 5/3$ and $\bar{m} = 0.613$ (\citeauthor{vasc..00}). In order to
obtain the ratio of protons to neutral atoms, we solve the Saha equation
for the values of the temperature and density given in Figure \ref{fig2}
and in equation (\ref{dens}), respectively. The obtained values are
shown in Figure \ref{fig3}. In solving the damping lengths for the
profiles described in \S \ref{magneto} (equations \ref{r} - \ref{B}
and Fig. \ref{fig2}), we use $r_m = r_{in} = 2.2 R_{\sun}$ and 0.74 rad
$\la \theta \la 1.57$ rad.  The results of the calculations are shown
in Figures \ref{fig4ab} and \ref{fig5ab}.

In Figure \ref{fig4ab}, we plot the heating rates for the adiabatic
heating mechanism (\citeauthor{martin}) (Fig. \ref{fig4ab}a) and
for the wave damping mechanisms (Fig. \ref{fig4ab}b), described by
equations \ref{Hnl}, \ref{HT}, \ref{Hcol} and \ref{Hvisres}. It can
be seen that the Alfvenic heating rates are larger than the adiabatic
heating rate. Figure \ref{fig5ab}a shows that the damping lengths are
small compared to the stellar radius. Thus, as soon as the waves are
generated, they are rapidly damped. We note that the nonlinear damping
length is much bigger than the other damping lengths.

In Figure \ref{fig5ab}b, we show the time-scales associated with each
of the damping processes. For comparison, we estimate the time-scale
associated with free-fall accretion:

\eq \label{tff}
t_{ff} \sim \frac{\Delta {\rm L}}{v_{ff}} \sim \frac{10^{12}}{3 \times 10^{7}} 
\sim 3.3 \times 10^4 {\rm s} \sim 9 \; {\rm hr}.
\feq
As can be noted, Alfv\'en waves are dissipated very quickly. Even
the slowest damping-time (nonlinear) is $10^4$ times shorter than the
free-fall time.

\section{Sources for Alfv\'en waves in magnetic accretion tubes} 
\label{sources}

\subsection{Generation of Alfv\'en waves at the star's surface}
\label{generation}

\citet{scheur..88} proposed that, when the fluid in the accretion column
shocks with the stellar surface, the resultant turbulence generates
Alfv\'en waves.  Thus, in a first approximation, we consider that the
waves are generated only at the surface of the star. However, taking into
account the results showed in Figure \ref{fig5ab}a, we note that the waves
are dissipated locally.  Therefore, unless thermal conduction is large,
Alfv\'en waves cannot be considered an efficient heating mechanism for
the tube, in this scenario.  In order to determine whether the damping of
Alfv\'en waves might be responsible for the increase of the temperature
in the tube in this model, we estimate the thermal conductivity near
the star.

For an incompressible fluid (with $\rho \sim$ constant), the change in
temperature with time is given by the thermal conduction equation,

\eq \label{thermequ}
\frac{\partial T}{\partial t} = -{\bf v} \cdot \nabla T + \frac{\kappa}
{\rho c_v} \nabla^2 T - \frac{1}{c_v} {\cal L}(\rho, T),
\feq
where $\kappa$ is the thermal conductivity coefficient, $c_v$ is the
specific heat at constant volume, and ${\cal L}(\rho, T) = \Gamma -
\Lambda$, the energy loss rate. Here, $\Gamma$ represents the energy
gains and $\Lambda$, the energy losses. In this work, we assume that
$\cal{L}$ is approximately zero. Thus, the energy gains are balanced by
energy losses \citep{shu92}.

In order to estimate the relative importance of thermal conduction in
magnetic flux tubes, we divide the first term by the second term on the
right hand side of equation (\ref{thermequ}). This gives

\eq \label{razao}
\frac{|-{\bf v} \cdot \nabla T|}{\frac{\kappa}{\rho c_v} \nabla^2 T}
\sim \left(\frac{\rho c_v}{\kappa}\right) v \Delta {\rm L}.
\feq

The specific heat at constant volume for a non relativistic, non degenerate,
monoatomic ideal gas is

\eq \label{cv}
\frac{c_v}{\cal{M}} = \frac{3 k}{2m},
\feq
where $\cal{M}$ is the molecular weight and $k$ is Boltzmann's constant.
>From \citet{spi62}, the thermal conductivity coefficient is

\eq \label{kapa}
\kappa \approx 2 \times 10^{-4}\frac{T^{5/2}}{Z^4 \, {\rm ln} \Lambda},
\feq
where $Z$ is the particle charge and ${\rm ln} \Lambda$ is the Coulomb 
logarithm.

Substituting equations (\ref{cv}) and (\ref{kapa}) in equation
(\ref{razao}), we obtain

\eq \label{razao2}
\frac{|-{\bf v} \cdot \nabla T|}{\frac{\kappa}{\rho c_v} \nabla^2 T}
\sim \frac{3 \, n \, k \times 10^4}{4} \frac{v \, \Delta {\rm L} \, {\rm  ln} 
\Lambda}{T^{5/2}}.
\feq
The $\Lambda$ parameter is given by

\eq \label{lamb}
\Lambda \sim 1.3 \times 10^4 \frac{T^{3/2}}{n_e^{1/2}},
\feq
where $n_e$ is the electron density. We assume free-fall velocities
on the order of 300 ${\rm km \; s}^{-1}$, den\-si\-ties and electron
densities of a\-bout $10^{12} \; {\rm cm}^{-3}$ and $5 \times 10^{11}
\; {\rm cm}^{-3}$, respectively, and temperatures around 8000 K
\citep{muz..98}. The length of the tube is approximately $10^{12}$
cm. Using these values in equation (\ref{razao2}), we obtain

\eq \label{razao3}
\frac{|-{\bf v} \cdot \nabla T|}{\frac{\kappa}{\rho c_v} \nabla^2 T}
\sim 5 \times 10^{10}.
\feq

Therefore, the term related to thermal conductivity, $(\kappa/ \rho c_v)
\nabla^2 T$, is very small, compared to the term $|-{\bf v} \cdot \nabla
T|$. We thus conclude that temperature gradients are maintained in the
tube because thermal conduction is not efficient. Estimating the time
scale related to this process, we obtain

\eq \label{tterm}
t_{therm} \sim \frac{\rho c_v}{\kappa} (\Delta {\rm L})^2 \sim 
5.5 \times 10^7 \; {\rm years}.
\feq

This time scale is very long, compared to the free fall time scale
($t_{ff} \sim 9$ hr). We conclude, therefore, that thermal conduction is
not important for the region near the star and, consequently, Alfv\'en
waves generated near the shock region cannot contribute to an increase
in temperature of the whole tube.

\subsection{Generation of Alfv\'en waves in the tube}

\citet[ and references therein]{jafe..90} proposed that the
Kelvin-Helmholtz (K-H) instability produced in an extragalactic jet by the
shear in the plasma flow with respect to the ambient medium can generate
MHD waves. They suggested that these waves, once damped, can give rise
to currents capable of generating magnetic fields. In the case of the
magnetic funnels of T Tauri stars, there is also a shear between the
gas falling onto the star and the external medium. This flow, however,
is sub-Alfv\'enic and, therefore, the development of Kelvin-Helmholtz
modes is not expected. In fact, \citet{hardee..92} performed an analytical
and numerical analysis of the Kelvin-Helmholtz instability in magnetized
jets and show that the fundamental mode is completely stabilized if the
magnetic Mach number [$M_{ms} = v_{tube} / \sqrt{c_{s}^2 + v_{A}^{2}}$]
is less than unity.  

\begin{figure}[htb!]
\epsscale{0.9}
\plotone{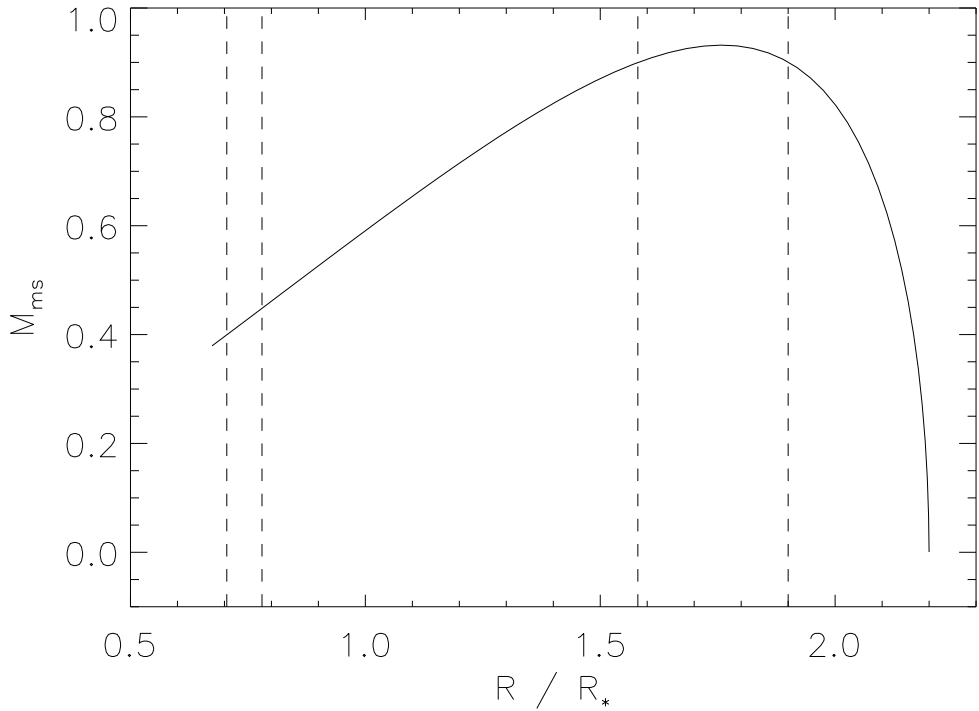}
\caption{{\small Magnetosonic Mach numbers for the tube. The values
varies from $5 \times 10^{-4}$ to $0.9$. The vertical dashed lines mark
where in the tube the conditions for the development of the K-H reflection
modes hold. There are two regions, one near the star where $0.705 \leq R /
R_{\ast} \leq 0.78$ and the other near the disk, where $1.58 \leq R / R_{\ast}
\leq 1.9$.} \label{machms}}
\end{figure}

However, they also find that for some magnetosonic mach numbers the
reflection mode is excited. There are two ranges shown in their Figure
1 where this can happen: $0.4 \leq M_{ms} \leq 0.45$ and $M_{ms}
\geq 0.9$. In Figure \ref{machms}, we plot magnetosonic Mach numbers
obtained for the parameters of the tube. We see that two regions can
support reflection modes of the K-H instability: one, near the star,
extends from $0.705 R_{\ast}$ to $0.780 R_{\ast}$. The other region,
near the disk, extends from $1.58 R_{\ast}$ to $1.90 R_{\ast}$, where
$M_{ms} \geq 0.9$. Thus, we see that Alfv\'en waves can be generated in
two regions in the tube.

We can also generate turbulence if the accretion is variable. In
this case, there will be shocks inside the tube, which can perturb the
environment. Indications for variable accretion have come from different
sources, for example, the internal knots observed in protostellar
jets \citep[for a recent review of both theory and observations of
Herbig-Haro jets, see][]{bo..99}. \citet{sano..99} made numerical
simulations of weakly ionized, magnetized accretion disks and found that
a magnetorotational instability \citep[for a review, see][]{balbus..98}
acts beyond 15 AU. The infalling matter accumulates in the region $\sim
15$ AU and a gravitational instability may be important in transporting
the flow to smaller radii. Even if a larger region of the disk may be
made susceptible to the action of the magnetorotational instability
due to ionization by cosmic rays \citep{gammie96} or X-rays from the
central star \citep{igea..99} or by the damping of Alfv\'en waves
(\citeauthor{vasc..00}), it is likely that there is some region where
the magnetorotational instability cannot occur \citep{gammie96} and a
gravitational instability probably occurs allowing accretion to smaller
radii.

There is no consensus about the position of the truncation radius,
$R_{trun}$, \citep{hart98}. The interaction of the star's magnetic field
with the accretion disk can be highly dynamic, so that, changes in the
position of $R_{trun}$ can occur \citep{safier98}. This would give rise
to variable accretion. \citet{johns..95}, analyzing the line profile
variability of SU Aurigae, argue that variation of the red absorption
feature in H$\beta$ is evidence for unsteady accretion.

The possibility of a turbulent component in accretion tubes has
been proposed by a number of people. \citet{basri90} suggested
that wings of Balmer lines are formed in a turbulent region near the
star. \citet{edw..94} suggested that Alfv\'en waves could be the source
for such a turbulence and \citet{johns..95} reproduced some symmetric
Balmer lines profiles by adding turbulent velocity components at the base
of an expanding wind.

Taking all this evidence into account, we suggest that Alfv\'en waves
are generated in the tube. In this paper we do not address the problem
related to the excitation of the Alfv\'en waves generated in a turbulent
medium. Instead, we use results from previous theoretical, solar wind
and laboratory findings.

A magnetized, perturbed environment excites MHD modes, in particular,
the Alfvenic mode. For example, laboratory plasma experiments
\citep*{moralez..98} show excitation of Alfv\'en modes with frequencies
around 0.1$\Omega_i$, where $\Omega_i$ is the ion-cyclotron frequency. The
frequency spectrum is seen to obey a Kolmogorov distribution. A frequency
spectrum of Alfv\'en waves generated by turbulence could be a Kolmogorov
one. Moreover, radio-wave scintillation observations of the ionized
interstellar medium show a density fluctuation spectrum that approximately
obeys a Kolmogorov distribution \citep*{armstrong..95,lith..01},
i.e. the fluctuations are approximately proportional to $\lambda^{1/3}$,
where lambda is the size of the density fluctuation. Recent theories
and simulations on MHD turbulence confirm the Kolmogorov spectrum for
Alfv\'en waves \citep{goldreich..95,goldreich..97,lith..01,maron..01}.
\citet{lith..01} extended the work of \citet{goldreich..95,goldreich..97}
which treated incompressible MHD turbulence, to include the effects
of compressibility and particle transport. They calculated small-scale
density spectra in turbulent interstellar plasmas. They also considered
compressible turbulence in plasmas with $\beta$ less than unity that match
those found in the magnetic tubes of T Tauri stars. They argued that
the dynamics of the cascade is roughly independent of $\beta$. In the
above mentioned papers, it was assumed that MHD wave-packets propagate
at the Alfv\'en speed, in a direction either parallel or antiparallel
with respect to the local mean magnetic field and that the nonlinear
interactions are restricted to collisions between oppositely directed
wave-packets. \citet{maron..01} made simulations of the interaction
between oppositely directed Alfv\'en waves.

In our previous paper (\citeauthor{vasc..00}), instead of a wave
spectrum, we assumed a mean frequency, $\varpi = F w_i$, while in
this paper, we treat different frequency values. The wave frequency
obtained from laboratory experiments \citep[e.g.,][]{moralez..98},
$\Omega_{sup} \sim 0.1 \Omega_i$, could be taken as the upper limit for
the wave spectrum. On the other hand, \citet{scheur..88} showed that
Alfv\'en wave frequencies can be greater than $10^{-5} \Omega_i$.  We,
therefore, study the frequency interval $10^{-5} \Omega_i \leq \Omega
\leq 0.1 \Omega_i$. Assuming this frequency spectrum, we calculate the
damping lengths and corresponding $f = \langle \delta v^2 \rangle^{1/2}
/ v_A$ parameter for all the damping mechanisms treated in section
\ref{dampmech}. Thus, hereafter, $f$ is not a free parameter anymore. We
assume that the energy released by the damping of the Alfv\'en waves in
a given volume is radiated away. Thus, we have

\eq \label{consist}
\int_{V} H_{nl, T} dV = \tau \sigma T^4 A,
\feq
where $T$ is given by Figure \ref{fig2}, $\tau$ is the optical depth,
$\sigma$ is the Stefan-Boltzmann constant, and $A$, the area. For a
given $\tau$ and temperature profile, we know the amount of energy that
is radiated away. This must be equal to the energy dissipated by the
Alfv\'en waves, which is determined by $f$ and the damping mechanism used.

\begin{figure}
%\epsscale{1.03}
\plotone{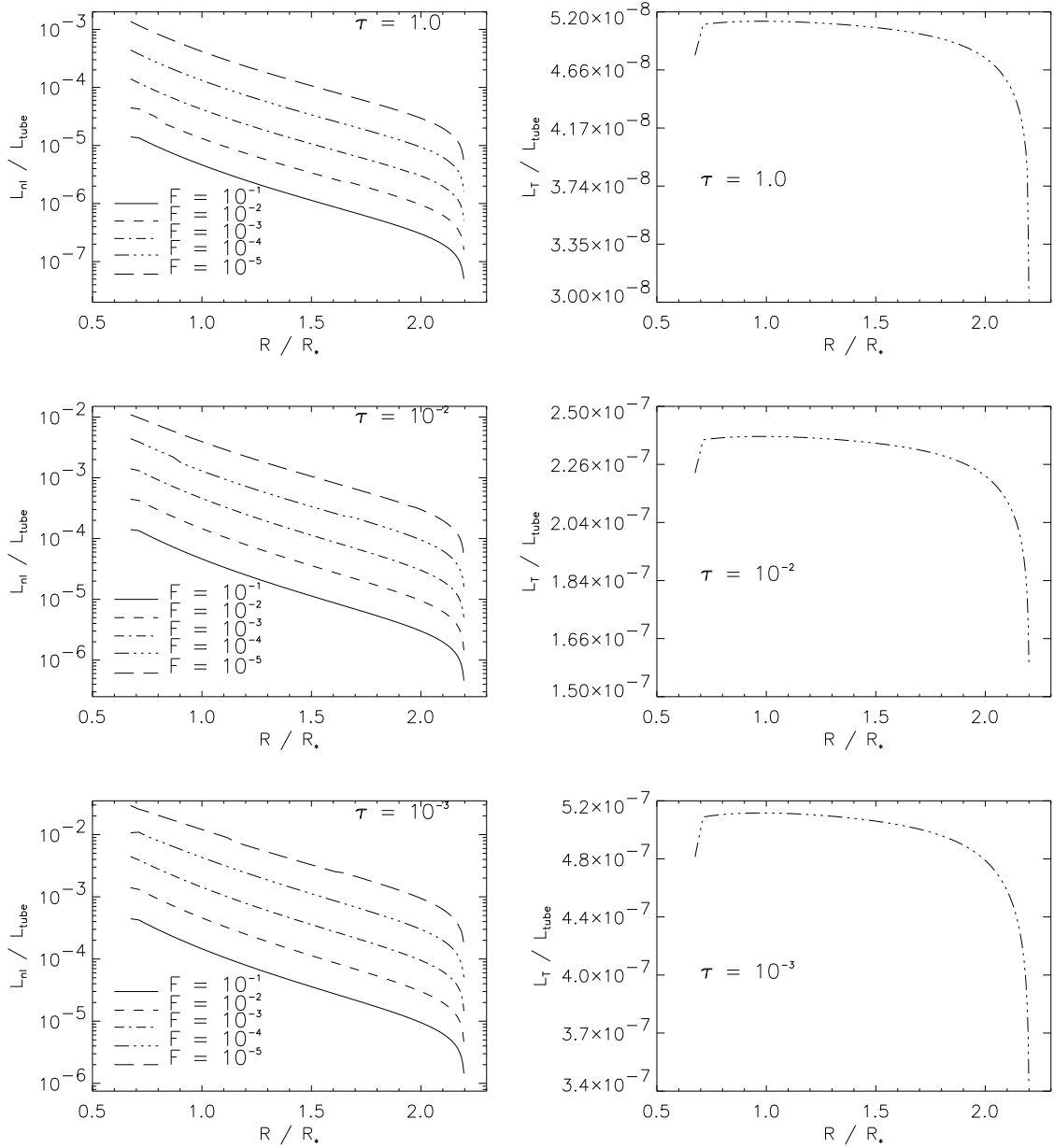}
\caption{{\small Variation of the damping lengths with tube
position. The damping lengths are divided by the tube length. Each
column shows the damping lengths related to one damping mechanism as a
function of the wave frequency and the optical depth. The left column
shows the nonlinear damping length; the right, the turbulent damping
length. Optical depth decreases from top to bottom ($\tau = 1.0, 10^{-2}$
and $10^{-3}$).} \label{fig7}}
\end{figure}

Figure \ref{fig7} shows the results of our calculations, which were made
for three different values of the optical depth $\tau$ and for $10^{-5}
\leq F \leq 0.1$. Since the star is not strongly obscured by the tube, it
is likely that the mean optical depth is less than 1. Taking into account
the values of the adiabatic heating \citep*{martin} and calculating the
associated optical depth, we obtain $\tau \sim 10^{-3}$. However, for
the sake of completeness, we made the calculations for $\tau = 10^{-2}$
and $\tau = 1$ as well. Figure \ref{fig7} shows the damping lengths
divided by the length of the tube, calculated according to equation
(\ref{consist}), for the nonlinear and turbulent damping mechanisms
mentioned above.

The left column of Figure \ref{fig7} shows the nonlinear damping
lengths. We note that even for $\tau = 10^{-3}$ with the lowest frequency
$(F = 10^{-5})$, the damping length is less than the tube length. This
implies, as seen in section \ref{generation}, that the nonlinear damped
waves cannot heat the whole tube if they are formed only at the surface
of the star. The same conclusion applies to turbulent damping (right
column of Figure \ref{fig7}). According to equations (\ref{gamanl})
and (\ref{consist}), the dependency of the nonlinear damping length
on the optical depth is $L_{nl} \propto \tau^{-1/2}$, the same as for
frequency, $L_{nl} \propto F^{-1/2}$. For the turbulent damping, there
is no dependency on frequency, whereas $L_{tur} \propto \tau^{-1/3}$
(equations \ref{Ltur} and \ref{consist}).

\begin{figure}
%\epsscale{1.05}
\plotone{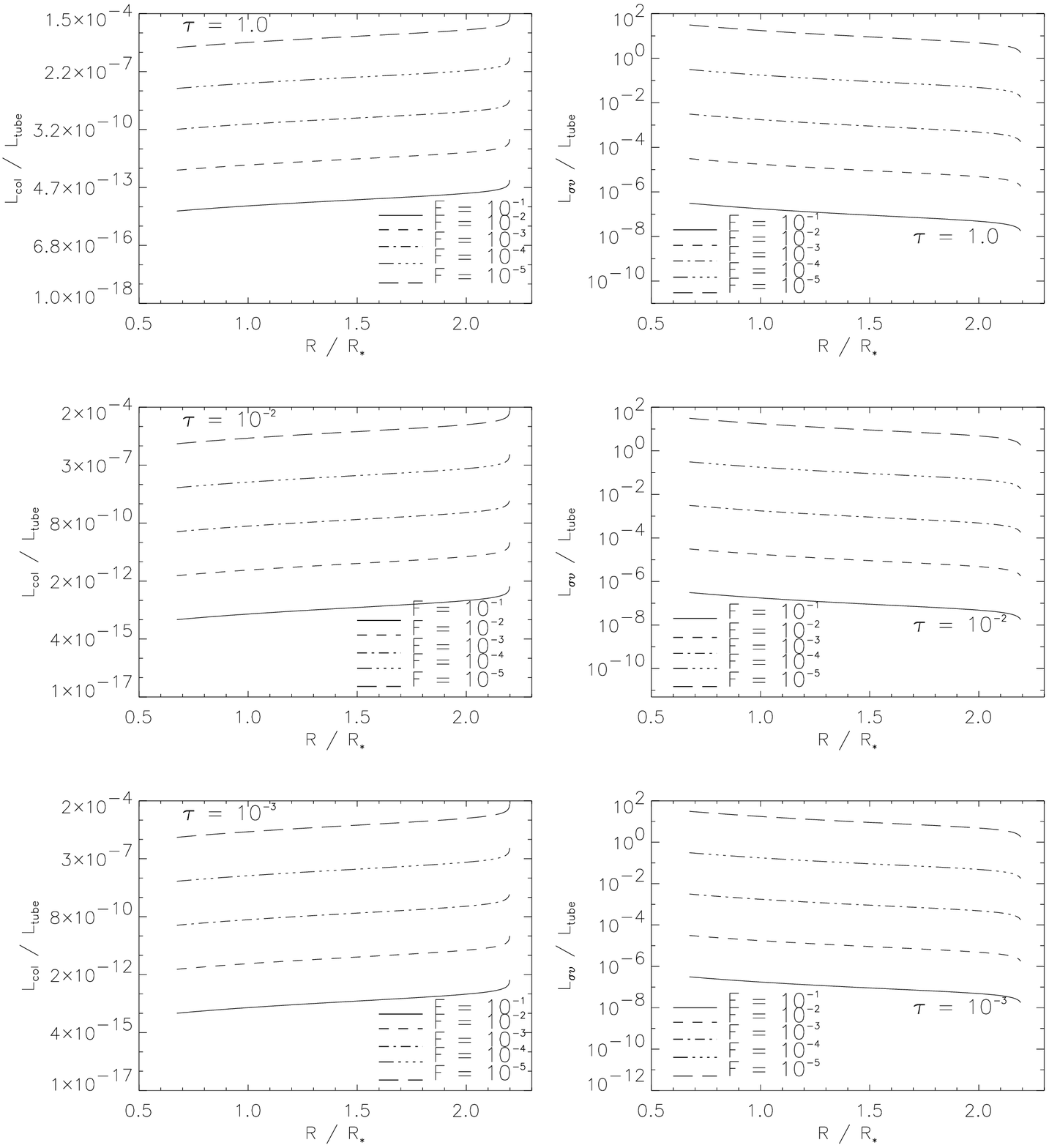}
\caption{{\small Damping lengths {\it versus} tube position as a
function of wave frequency and optical depth. The damping lengths are
divided by the tube length. In each pannel, different lines correspond to
different wave frequencies, that range from $0.1 \Omega_i$ to $10^{-5}
\Omega_i$. The results for waves undergoing collisional damping are
plotted in the left column. Viscous-resistive damping lengths are
showed in the right column. Optical depths decreases from top to bottom.}
\label{fig8}}
\end{figure}

Figure \ref{fig8} shows the damping lengths for the collisional and
viscous-resistive damping, divided by the tube length. The collisional
damping lengths are plotted in the left column and, in the right column,
we plot the results for the viscous-resistive mechanism. As noted for
the nonlinear and turbulent cases, damping lengths for the collisional
mechanism are smaller than the tube length, for all frequencies
considered. Thus, waves formed at the surface of the star undergoing
collisional damping cannot heat the whole tube. The only mechanism
that might be able to heat the tube entirely is viscous-resistive
damping. If the wave frequency is lower than $10^{-4} \Omega_i$, the
viscous-resistive damping length is comparable to the tube length,
regardless of the optical depth. We note that for collisional damping,
the variation of the damping length with frequency is very large, with
$L_{coll}$ ranging from $\approx 10^{-14}$ for $F = 0.1$ to $\approx
10^{-6}$, for $F = 10^{-5}$, near the star, although variations with
optical depth are almost negligible. In this case, a slight variation
in frequency causes a large variation in the damping length. However,
it is necessary to have frequencies well bellow $10^{-5} \Omega_i$
in order that the damping length be comparable to the length of the tube.

For viscous-resistive mechanism, the damping length also varies greatly
with frequency (about 7 orders of magnitude near the star) and again the
dependency with optical depth is weak. For waves with frequencies less
than $10^{-4} \Omega_i$, the damping length is approximately equal the
length of the tube.

Thus, we note a great variation of damping lengths with wave frequency
for all mechanisms, except for the case of turbulent damping. In general,
the lower the frequency, the larger the damping lengths.

\begin{figure}
\epsscale{1.05}
\plotone{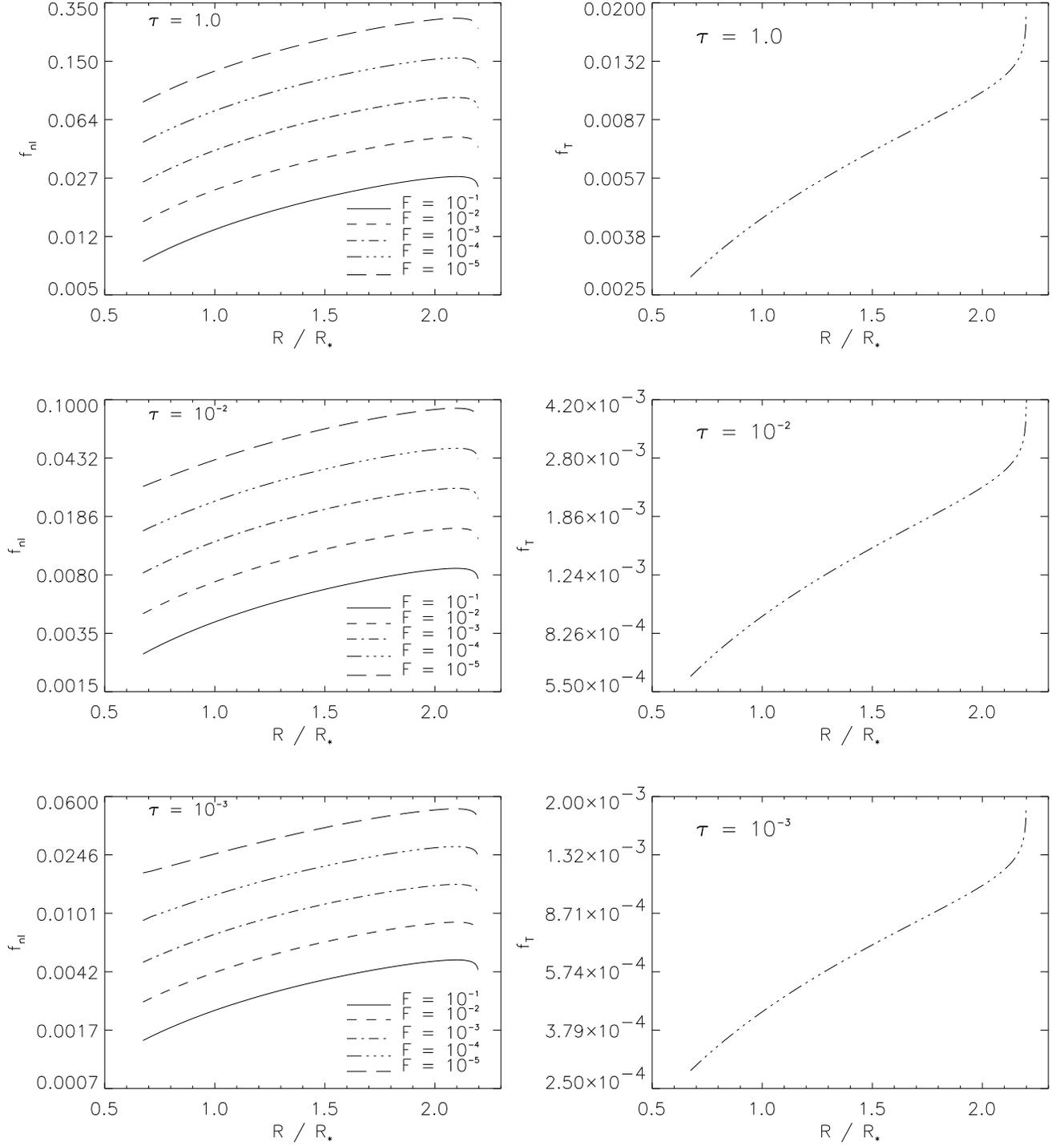}
\caption{{\small Degree of turbulence $f \equiv \langle \delta
v^2 \rangle^{1/2} /v_A$. Each pannel correspond to a different optical
medium.  Also, different lines in the pannels correspond to different
wave frequencies.  The column sequency is: nonlinear damping (left)
and turbulent damping (right).} \label{fig9}}
\end{figure}

\begin{figure}
\epsscale{0.9}
\plotone{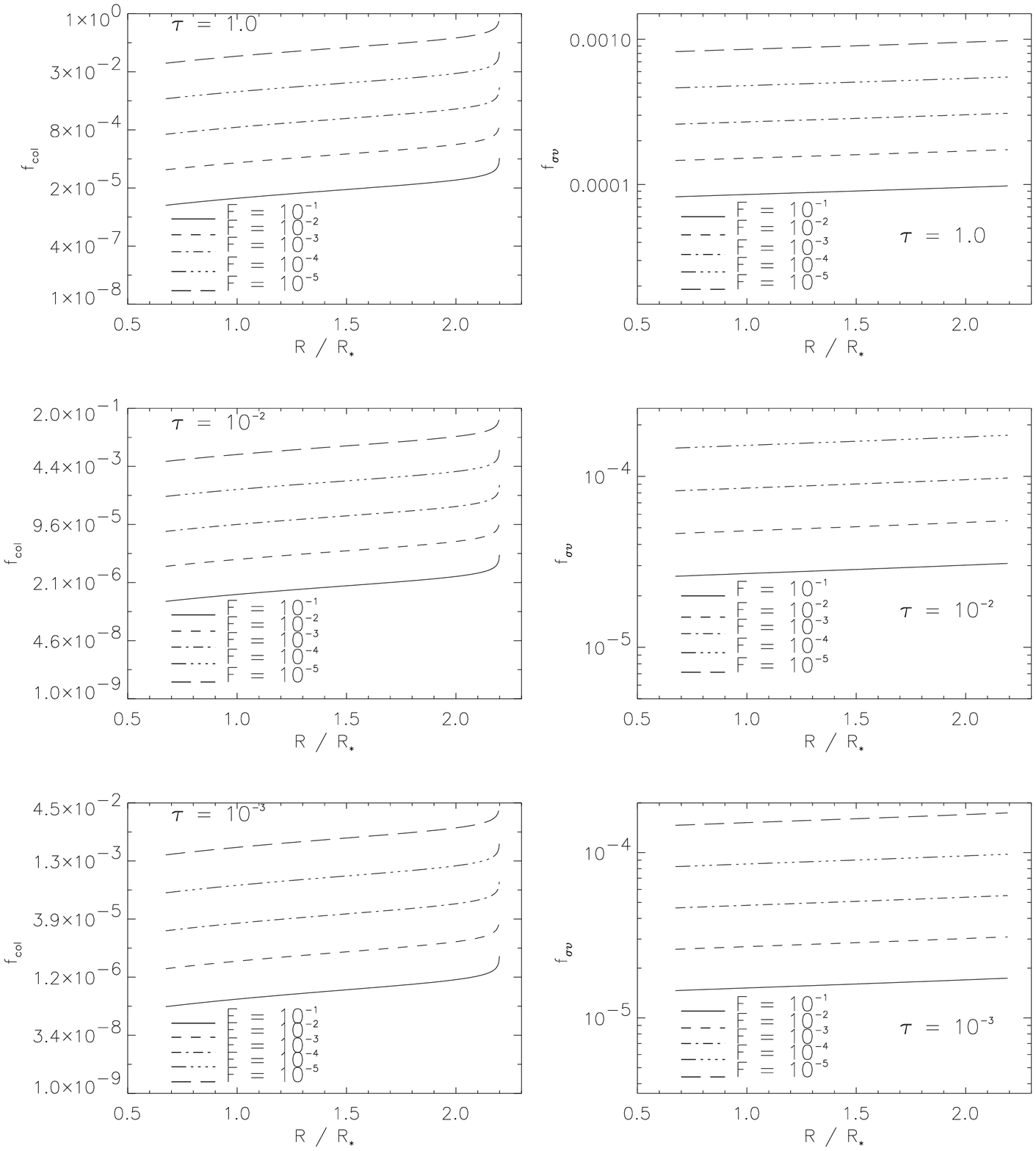}
\caption{{\small Parameter $f \equiv \langle \delta v^2
\rangle^{1/2} /v_A$. Each pannel correspond to a different optical
medium.  Also, different lines in the pannels correspond to different
wave frequencies. In the left column are showed the results obtained
for collisional damping mechanism and in the right column, the results
for viscous-resistive damping.} \label{fig10}}
\end{figure}

Figure \ref{fig9} shows the $f$ parameter for nonlinear (left column)
and turbulent damping (right column) mechanisms. In Figure \ref{fig10},
the calculations are for collisional (left column) and viscous-resistive
damping (right column). The same analysis previously made for the damping
lengths also applies here. In general, in order to fit the observations,
a higher degree of turbulence is necessary if the gas is optically thick,
even for the collisional and viscous-resistive mechanisms, although for
the nonlinear and turbulent damping the dependency with optical depth
is stronger. Again, the value of $f$ varies greatly with frequency.

\begin{figure}
%\epsscale{1.05}
\plotone{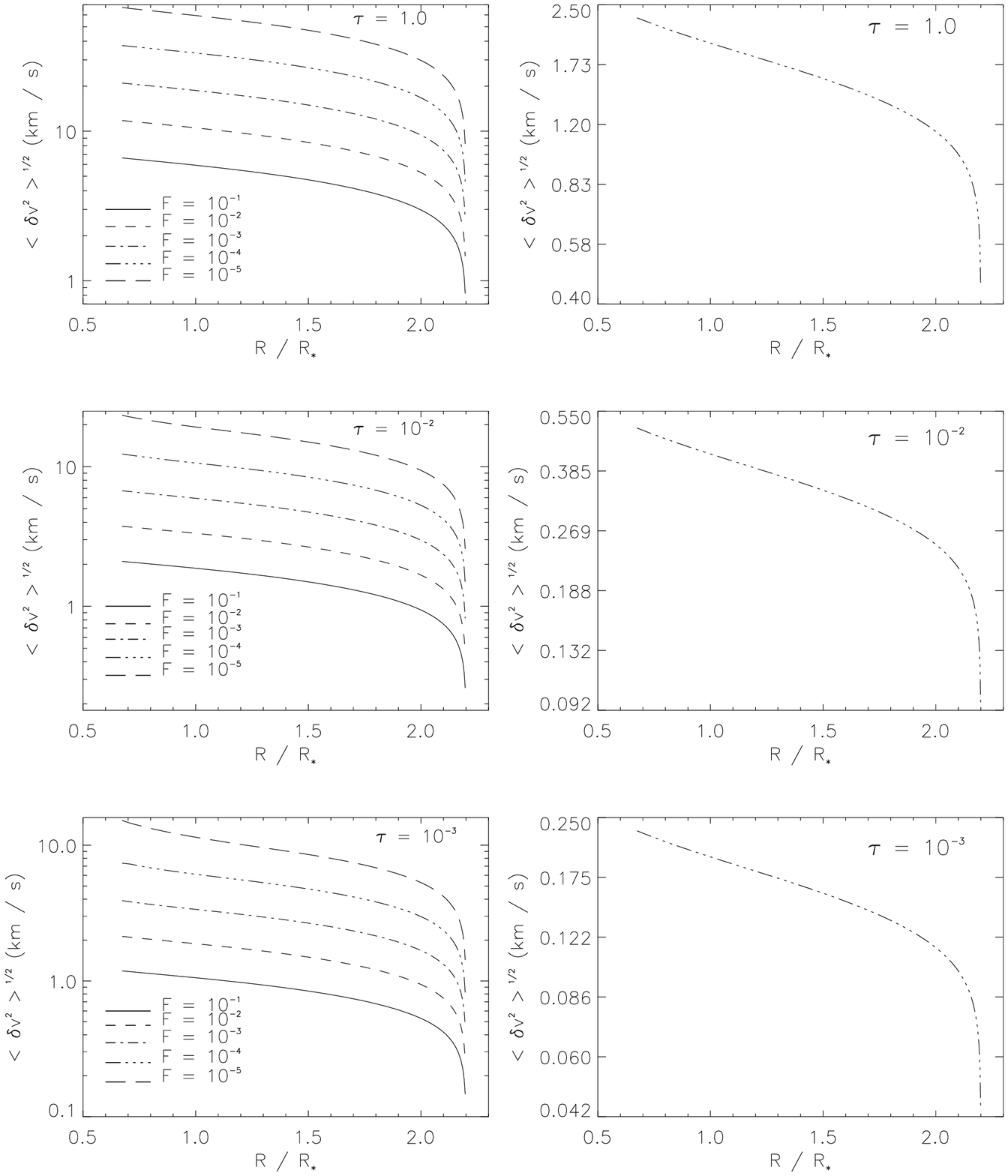}
\caption{{\small The root mean square velocity $\langle \delta v^2
\rangle^{1/2}$, given in kilometers per second. Each pannel correspond
to a different optical medium.  Also, different lines in the pannels
correspond to different wave frequencies. Nonlinear damping: left column;
turbulent damping: right column.} \label{fig11}}
\end{figure}

\begin{figure}
%\epsscale{1.1}
\plotone{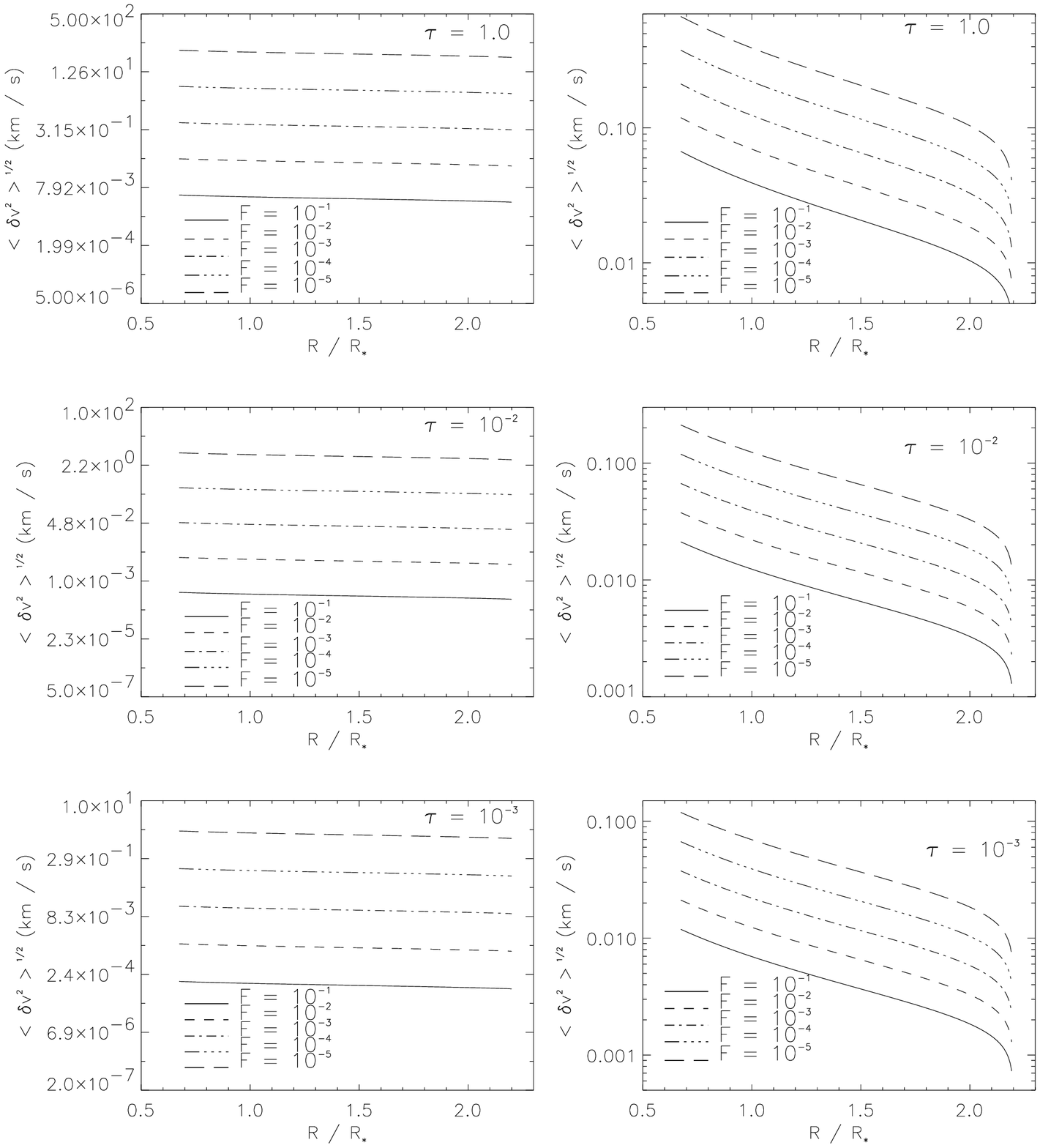}
\caption{{\small The root mean square velocity $\langle \delta v^2
\rangle^{1/2}$, given in kilometers per second. Each pannel correspond
to a different optical medium.  Also, different lines in the pannels
correspond to different wave frequencies. Velocities obtained using
the collisional mechanism are showed in the left column; velocities
obtained for viscous-resistive damping are plotted in the right column.}
\label{fig12}}
\end{figure}

We can also analyze the values of the root mean square velocity, $\langle
\delta v^2 \rangle ^{1/2}$ which, according to equation (\ref{dv2})
can be given in terms of the parameter $f$ and the Alfv\'en speed. The
results are shown in Figures \ref{fig11} and \ref{fig12}, where the rms 
velocity is given in kilometers per second.

The left column of Figure \ref{fig11} shows $\langle \delta v^2
\rangle^{1/2}$ for the nonlinear case, for three different optical
depths. The value of $\langle \delta v^2 \rangle^{1/2}$ for the
turbulent damping is shown in the right column of Figure \ref{fig11}. In
Figure \ref{fig12} we plot the rms velocity for the collisional and
viscous-resistive mechanisms in the left and in the right columns,
respectively. We find that the greater the optical depth, the greater
the rms velocity, which increases with decreasing wave frequency.
For the nonlinear and collisional mechanisms, the rms velocity can be
as great as 60 km s$^{-1}$ for an optically thick medium and a very low
wave frequency ($F = 10^{-5}$). The $\langle \delta v^2 \rangle^{1/2}$
is smallest for the case of the viscous-resistive mechanism.

\section{Conclusions} \label{conc}

In this work, we evaluated the role of damped Alfv\'en waves in the heating of
the tubes in magnetospheric accretion models. We analyzed four damping
mechanisms: 1) nonlinear; 2) turbulent; 3) viscous-resistive; and 4)
collisional.

Using the values f = 0.004 and F = 0.1, all the damping lengths
are found to be small compared with the length of the tube. Since our
calculations show that the thermal conductivity of the gas is very small,
these high-frequency waves, if generated only at the surface of the star,
cannot heat the whole tube.

Extending the frequency range of the Alfv\'en waves from F = 0.1 to
$10^{-5}$, we found that the damping lengths varied strongly with frequency,
spanning eight orders of magnitude for collisional damping. The parameter $f$
was eliminated, imposing the constraint that the Alfvenic heating was
converted into radiation in different optical media. The damping lengths were
found to be much smaller than the tube length, except when the lowest frequency
waves had viscous-resistive damping. In this case, the damping length was
greater than the tube. If this were the dominant mechanism, Alfv\'en waves 
generated at the surface of the star could heat the whole tube. For the other 
cases, waves produced locally are required to explain the observations.
The degree of turbulence $\langle \delta v^2 \rangle^{1/2}$ necessary
to heat the tube was constrained, taking into account the optical depth of 
the gas. The larger the optical depth, the greater the degree of turbulence
required. We found that the turbulent velocity of the Alfv\'en waves can 
reach 100 km s$^{-1}$, for very low frequency waves ($F = 10^{-5}$) undergoing
 collisional damping.

\acknowledgments The authors thank an anonymous referee for useful suggestions. MJV would like to thank Adriano H. Cerqueira, Silvia
H. P. Alencar and Bruno V. Castilho for very useful discussions and
suggestions. MJV also thanks the Brazilian agency FAPESP (Proc. No. 
96/00677-3) and PROPP/UESC for finantial support. VJP and RO thank the
federal Brazilian agency CNPq for partial support. The authors would
also like to thank the project PRONEX (41.96.0908.00) for partial
support.

\begin{deluxetable}{cc}
\tablewidth{0pt}
\tablecaption{Parameters of the magnetospheric model. \label{tab1}}
\tablehead{
\colhead{Parameter} & \colhead{Value}}
\startdata
$M_{\ast}$ & $0.8 \; M_{\sun}$ \\
$R_{\ast}$ & $2 \; R_{\sun}$ \\
$\dot{M}$ & $10^{-7} \; M_{\sun} \; yr^{-1}$ \\
$r_{in}$ & $2.2 \; R_{\sun}$ \\
$r_{out}$ & $3 \; R_{\sun}$ \\
$m$\tablenotemark{a} & $4.38 \times 10^{36}$ erg ${\rm G}^{-1}$ \\
\enddata
\tablenotetext{a}{This value was obtained by assigning a value of 1 kG to
the magnetic field at the surface of the star.}
\end{deluxetable}

%\begin{deluxetable}{ccccccc}
%\tablewidth{0pt}
%\tablecaption{Parameters of the Kelvin-Helmholtz instability criterium. 
%\label{tab2}}
%\tablehead{
%\colhead{R} & \colhead{$\theta$} & \colhead{Mach number} & \colhead{$\eta$}
%& \colhead{$(M_T/2)^{3.3}$} & \colhead{Tube radius (cm)} & 
%\colhead{$\lambda_{K-H}$ (cm)}}
%\startdata
%0.67 & 0.74 & 29.5 & 2165.1 & 7170.2 & $5.059 \times 10^{10}$ &
%$1.491 \times 10^{12}$ \\
%0.83 & 0.81 & 27.7 & 1195.9 & 5811.5 & $5.798 \times 10^{10}$ &
%$1.603 \times 10^{12}$ \\
%1.07 & 0.91 & 24.7 & 5712.4 & 4005.6 & $6.896 \times 10^{10}$ &
%$1.703 \times 10^{12}$ \\
%1.33 & 1.01 & 21.5 & 3590.1 & 2512.8 & $7.941 \times 10^{10}$ &
%$1.703 \times 10^{12}$ \\
%1.57 & 1.11 & 17.8 & 3111.2 & 1363.7 & $8.891 \times 10^{10}$ &
%$1.585 \times 10^{12}$ \\
%1.79 & 1.21 & 13.8 & 3653.9 & 5842.8 & $9.710 \times 10^{10}$ &
%$1.339 \times 10^{12}$ \\
%1.98 & 1.31 & 9.4  & 5513.3 & 1658.7 & $1.037 \times 10^{11}$ &
%$9.756 \times 10^{11}$ \\
%2.11 & 1.40 & 5.0  & 1055.7 & 20.90  & $1.083 \times 10^{11}$ &
%$5.442 \times 10^{11}$ \\
%2.19 & 1.50 & 1.3  & 3397.0 & 0.27   & $1.109 \times 10^{11}$ &
%$1.482 \times 10^{11}$ \\
%\enddata
%\end{deluxetable}

\end{document}